\documentclass[12pt]{article}
\usepackage{amsmath,amssymb,amscd,amsbsy,amsgen,amsopn,amstext,
amsxtra}

\usepackage[dvips]{graphicx}

\begin{document}

\begin{flushright}
\end{flushright}

\vskip 0.5 truecm

\begin{center}
{\Large{\bf A universally valid Heisenberg uncertainty relation}}
\end{center}
\vskip .5 truecm
\centerline{\bf  Kazuo Fujikawa}
\vskip .4 truecm
\centerline {\it Mathematical Physics Laboratory, RIKEN 
Nishina Center}
\centerline {\it Wako, Saitama 351-0198, 
Japan}
\vskip 0.5 truecm

\makeatletter
\makeatother

\begin{abstract}
A universally valid Heisenberg uncertainty relation is proposed by combining the universally valid error-disturbance uncertainty relation of Ozawa with the relation of Robertson. This form of the uncertainty relation, which is defined with the same mathematical rigor as the relations of Kennard and Robertson, incorporates both of the intrinsic quantum fluctuations and measurement effects.  
\end{abstract}

\section{Heisenberg uncertainty relation}   
The uncertainty relation forms the basis of entire quantum theory~\cite{heisenberg, kennard, robertson}. The original formulation by Heisenberg~\cite{heisenberg} is based on a gedanken experiment and emphasizes  measurement processes.
On the other hand, the formulations by Kennard~\cite{kennard} and Robertson~\cite{robertson}, which are mathematically well-defined, evaluate only the intrinsic quantum fluctuations without any reference to measurement processes. The formulations by Kennard and Robertson are widely accepted as the mathematical expression of the Heisenberg uncertainty relation. In the spirit of the original formulation of Heisenberg, however, it may be desirable to incorporate the effects of measurement in the formulation of the uncertainty relation. It may be desirable to express the uncertainty relation in terms of the measurement error $\epsilon(A)$ of the variable $A$ and the disturbance $\eta(B)$ in the conjugate variable $B$, for example.  From this point of view, the recent beautiful experiment~\cite{hasegawa}
provides a strong support for the idea of a universally valid error-disturbance uncertainty relation proposed by Ozawa~\cite{ozawa1,ozawa2,ozawa3,ozawa4} which is 
written as
\begin{eqnarray}
\epsilon(A)\eta(B)+\sigma(A)\eta(B)+\epsilon(A)\sigma(B)\geq \frac{1}{2}|\langle [A,B]\rangle|
\end{eqnarray}
where $\sigma(A)$ stands for the standard deviation.
This relation is derived from the identity
\begin{eqnarray}
[M^{out},B^{out}]&=&[M^{out}-A^{in},B^{out}-B^{in}]+[A^{in},B^{out}-B^{in}]\nonumber\\
&+&[M^{out}-A^{in}, B^{in}] + [A^{in}, B^{in}]
\end{eqnarray}
and the commutator of independent non-conjugate variables $[M^{out},B^{out}]=0$ combined with the triangle inequality, namely, $|\langle[M^{out}-A^{in},B^{out}-B^{in}] \rangle|+|\langle[A^{in},B^{out}-B^{in}] \rangle|+|\langle[M^{out}-A^{in}, B^{in}] \rangle|\geq |\langle[A^{in}, B^{in}] \rangle|$ and applying the  Robertson's relation for generic variables $A$ and $B$~\cite{robertson}
\begin{eqnarray}
\sigma(A)\sigma(B)\geq \frac{1}{2}|\langle [A,B]\rangle|
\end{eqnarray}
to each term on the left-hand side. The crucial property is that we do not impose any constraint on state vectors except for the positive metric Hilbert space and use only the natural commutator algebra in the derivation of (1), and thus the resulting relation is "universally valid". 

In the relation (1), $\sigma(A)=\sigma(A^{in})$ stands for the standard deviation $\sigma(A)=\langle (A-\langle A\rangle)^{2}\rangle^{1/2}$, and $\epsilon(A)$ stands for the measurement "error" of $A$ defined by 
\begin{eqnarray}
\epsilon(A)=\langle (M^{out}-A^{in})^{2}\rangle^{1/2}
\end{eqnarray}
where $A^{in}$ stands for the variable $A$  before the measurement and $M^{out}$ stands for the meter variable after the measurement of $A$. The quantity 
 $\eta(B)$ stands for the "disturbance" of the conjugate variable
 $B$ defined by
\begin{eqnarray}
\eta(B)=\langle (B^{out}-B^{in})^{2}\rangle^{1/2}
\end{eqnarray}
in terms of the variable $B$ which assumes $B^{in}$  before and $B^{out}$  after the measurement of $A$, respectively. We note that $\epsilon(A)\geq\sigma(M^{out}-A^{in})$ and $\eta(B)\geq\sigma(B^{out}-B^{in})$ which are used in the derivation of (1), for example, $\epsilon(A)\sigma(B)= \epsilon(A)\sigma(B^{in})\geq\sigma(M^{out}-A^{in})\sigma(B^{in})$. Following Ref.~\cite{ozawa1}, we often use $A$ and $B$ in place of initial variables $A^{in}$ and $B^{in}$, respectively, in the relation such as (1).

In this notation, Ozawa proposes~\cite{ozawa1} to identify the Heisenberg's original uncertainty relation with
\begin{eqnarray}
\epsilon(A)\eta(B)\geq \frac{1}{2}|\langle [A,B]\rangle|
\end{eqnarray}
which does not always hold~\cite{hasegawa, ozawa4}. This relation is not universally valid since we threw away two terms  on the left-hand side in the relation (1). In comparison, the universally valid Robertson's relation is   
\begin{eqnarray}
\sigma(M^{out}-A^{in})\sigma(B^{out}-B^{in})\geq \frac{1}{2}|\langle [M^{out}-A^{in},B^{out}-B^{in}]\rangle|
\end{eqnarray}
which may be re-written in the present notation of $\epsilon(A)$ and $\eta(B)$ as
\begin{eqnarray}
\epsilon(A)\eta(B)\geq \frac{1}{2}|\langle [M^{out}-A^{in},B^{out}-B^{in}]\rangle|.
\end{eqnarray}
Note that in general 
\begin{eqnarray}
|\langle [M^{out}-A^{in},B^{out}-B^{in}]\rangle|\neq |\langle [A^{in},B^{in}]\rangle|= |\langle [A,B]\rangle|
\end{eqnarray}
without any constraint on state vectors. See also (18) below.
Although each term in (1) has a definite meaning, the physical meaning of the entire relation (1) itself is less clear compared with (6).

We here suggest to combine the relation (1) with the standard Robertson's relation (3) in the form
\begin{eqnarray}
\bar{\epsilon}(A)\bar{\eta}(B)\geq |\langle [A,B]\rangle|
\end{eqnarray}
where
\begin{eqnarray}
\bar{\epsilon}(A)&\equiv&\epsilon(A)+\sigma(A)\nonumber\\
&=&\langle (M^{out}-A^{in})^{2}\rangle^{1/2}+\langle (A^{in}-\langle A^{in}\rangle)^{2}\rangle^{1/2},\nonumber\\
\bar{\eta}(B)&\equiv&\eta(B)+\sigma(B)\nonumber\\
&=&\langle (B^{out}-B^{in})^{2}\rangle^{1/2}+\langle (B^{in}-\langle B^{in}\rangle)^{2}\rangle^{1/2}.
\end{eqnarray} 
The relation (10) is similar to the Arthurs-Kelly relation~\cite{arthurs,arthurs2} and, as is explained later, it amounts to analyzing the {\em repeated measurements} of similarly prepared samples in quantum mechanics~\cite{note}. We emphasize that the relation (10) is based on the positive metric Hilbert space and the natural commutator algebra, and thus its validity is at the same level as the standard Kennard~\cite{kennard} and Robertson~\cite{robertson} relations.

Our proposal is to identify the relation (10) as a universally valid Heisenberg uncertainty relation which incorporates measurement effects. Physically, we identify 
$\bar{\epsilon}(A)$ in (11) as the "inaccuracy" in the measured values of $A$ and 
$\bar{\eta}(B)$ in (11) as the "fluctuation" of the conjugate variable $B$ after the measurement of $A$. The fact that in the relation (1) or (10) we first measure $A$ and then examine the variation in $B$ immediately after the measurement of $A$ (and also its experimental realization~\cite{hasegawa}) implies
the successive measurements rather than simultaneous measurements, and thus the appearance of the standard deviations is natural.

In the following, we would like to provide the motivation for writing the relation (10) and its physical interpretation: 
We start with the analysis of the "precise" measurement\ $\epsilon(A)=\langle (M^{out}-A^{in})^{2}\rangle^{1/2}=0$~\cite{ozawa1}, which suggests
\begin{eqnarray}
M^{out}|\psi\rangle\otimes|\xi\rangle=A^{in}|\psi\rangle\otimes|\xi\rangle
\end{eqnarray}
where $|\psi\rangle$ stands for the state vector of the system and $|\xi\rangle$ is a suitable state vector of the measuring apparatus~\cite{ozawa4, neumann}. Note that we are working in the Heisenberg representation, and the initial variables $A^{in}$ and $M^{in}$ act on the states $|\psi\rangle$ and $|\xi\rangle$, respectively, but $M^{out}$ acts on the state $|\psi\rangle\otimes|\xi\rangle$ due to the interaction term in the total Hamiltonian which drives $M^{in}$ to $M^{out}$.
We thus have the relation between standard deviations
\begin{eqnarray}
\sigma(M^{out})=\sigma(A^{in})=\sigma(A)
\end{eqnarray}
defined for the state $|\psi\rangle\otimes|\xi\rangle$,
namely, even for the "precise" measurement we have the standard deviation $\sigma(M^{out})=\sigma(A)$ of the measurement apparatus if one performs the {\em repeated} measurements in quantum mechanics. This property
is consistent with the identification of $\bar{\epsilon}(A)$ in (11) as the "inaccuracy" in the measured values of $A$ even for the  "precise" measurement. Similarly, one can identify $\bar{\eta}(B)$ in (11) as the inevitable "fluctuation" of the conjugate variable $B$ after the measurement of $A$, if one repeats the measurements of similarly prepared samples; even if $\eta(B)=\langle (B^{out}-B^{in})^{2}\rangle^{1/2}=0$, namely, for $B^{out}|\psi\rangle\otimes|\xi\rangle=B^{in}|\psi\rangle\otimes|\xi\rangle
$, we have the fluctuation $\sigma(B^{out})=\sigma(B^{in})=\sigma(B)$ in the variable $B^{out}$.

Our suggested relation (10) is similar to the Arthurs-Kelly relation~\cite{arthurs, arthurs2}. In the spirit of the original Arthurs-Kelly relation, one may consider $M^{out}=(M^{out}-A^{in})+A^{in}$ and 
\begin{eqnarray}
\sigma^{2}(M^{out})&=&\langle (M^{out}-A^{in})^{2}\rangle+
\langle A^{in}(M^{out}-A^{in})\rangle\nonumber\\
&&+\langle (M^{out}-A^{in})A^{in}\rangle+\sigma^{2}(A^{in})\nonumber\\
&=&\langle (M^{out}-A^{in})^{2}\rangle+\sigma^{2}(A^{in})
\end{eqnarray}
for the {\em unbiased} measurement which is assumed to satisfy $\langle (M^{out}-A^{in})\rangle=\langle A^{in}(M^{out}-A^{in})\rangle=\langle (M^{out}-A^{in})A^{in}\rangle=0$ for all $|\psi\rangle$ in $|\psi\rangle\otimes|\xi\rangle$. See, for example, Appendix in~\cite{appleby}. Similarly, for the unbiased disturbance $\langle (B^{out}-B^{in})\rangle=\langle B^{in}(B^{out}-B^{in})\rangle=\langle (B^{out}-B^{in})B^{in}\rangle=0$, one has
\begin{eqnarray}
\sigma^{2}(B^{out})
&=&\langle (B^{out}-B^{in})^{2}\rangle+\sigma^{2}(B^{in}).
\end{eqnarray}
One may thus define a direct analogue of the Arthurs-Kelly relation
\begin{eqnarray}
\sigma^{2}(M^{out})\sigma^{2}(B^{out})
&=&\{\langle (M^{out}-A^{in})^{2}\rangle+\sigma^{2}(A^{in})\}\nonumber\\
&&\times\{\langle (B^{out}-B^{in})^{2}\rangle+\sigma^{2}(B^{in})\}\nonumber\\
&\geq&\frac{1}{4}|\langle [A^{in},B^{in}]\rangle|^{2}\{
\frac{1}{\langle (B^{out}-B^{in})^{2}\rangle}+\frac{1}{\sigma^{2}(B^{in})}\}\nonumber\\
&&\times\{\langle (B^{out}-B^{in})^{2}\rangle+\sigma^{2}(B^{in})\}\nonumber\\
&\geq& |\langle [A^{in},B^{in}]\rangle|^{2}
\end{eqnarray}
by assuming
\begin{eqnarray}
&&\langle (M^{out}-A^{in})^{2}\rangle\langle (B^{out}-B^{in})^{2}\rangle\geq \frac{1}{4}|\langle [A^{in},B^{in}]\rangle|^{2},\nonumber\\
&&\sigma^{2}(A^{in})\sigma^{2}(B^{in})\geq \frac{1}{4}|\langle [A^{in},B^{in}]\rangle|^{2}.
\end{eqnarray}
The weakness of (16) is that one can show that the left-hand side of the first relation in (17) becomes 
\begin{eqnarray}
\langle (M^{out}-A^{in})^{2}\rangle\langle (B^{out}-B^{in})^{2}\rangle = 0
\end{eqnarray}
for the bounded operator $B$ such as the spin variable  for the "precise" measurement $\langle (M^{out}-A^{in})^{2}\rangle=0$ regardless of the value of $\frac{1}{4}|\langle [A^{in},B^{in}]\rangle|^{2}$ and thus the first relation in (17), which is equivalent to (6), does not hold in general as was demonstrated in
Ref.~\cite{ozawa4}. See also the recent experimental result~\cite{hasegawa}. One cannot maintain the direct analogue of the Arthurs-Kelly relation (16) in the present context. We need to consider (10) and (11) instead, which also incorporate both of the intrinsic quantum fluctuations and measurement effects. We note the inequality
\begin{eqnarray}
\bar{\epsilon}(A)\bar{\eta}(B)&=&\{\epsilon(A)+\sigma(A)\}
\{\eta(B)+\sigma(B)\}\nonumber\\
&\geq&\{\langle (M^{out}-A^{in})^{2}\rangle+\sigma^{2}(A^{in})\}^{1/2}\nonumber\\
&&\times\{\langle (B^{out}-B^{in})^{2}\rangle+\sigma^{2}(B^{in})\}^{1/2}\nonumber\\
&=&\sigma(M^{out})\sigma(B^{out}),
\end{eqnarray}
and in fact our analysis shows that  the equality sign does not hold in general. It may be worth noting that while the last inequality in (19) is valid if one assumes the unbiased measurement and disturbance, no such assumption is needed for the validity of (10).
 
\section{Conclusion} 
We suggested to exploit the certain freedom in the identification of the original  Heisenberg uncertainty relation which incorporates measurement effects.
The identification of (6) with the original Heisenberg uncertainty relation~\cite{ozawa1} emphasizes the direct effects of the measurement, but the uncertainty relation in this form does not always hold~\cite{hasegawa}. On the other hand, our identification of (10) with a universally valid  Heisenberg uncertainty relation emphasizes what we actually learn from   repeated measurements. Our identification has an advantage
in factoring the uncertainty relation into two factors with clear physical meanings, the inaccuracy $\bar{\epsilon}(A)$ in the measured values of the variable $A$ and the resulting fluctuation $\bar{\eta}(B)$ in the conjugate variable $B$. It is significant that our relation (10) can be proven with the same mathematical rigor as the relations of Kennard~\cite{kennard} and Robertson~\cite{robertson} without making the assumptions such as unbiased measurement. 
Our relation will be useful in the future applications of the universally valid error-disturbance uncertainty relation~\cite{ozawa1}.


\begin{thebibliography}{99}
\bibitem{note}
One may equally use the definitions,
$\bar{\epsilon}(A)\equiv\sigma(M^{out}-A^{in})+\sigma(A)$ and
$\bar{\eta}(B)\equiv\sigma(B^{out}-B^{in})+\sigma(B)$,
for which the relation (10) is still universally valid. In this connection, one may define $\bar{\epsilon}(B)\equiv\sigma(N^{out}-B^{in})+\sigma(B)$ where $N^{out}$ stands for the meter variable to measure the variable $B^{in}$ with $[M^{out},N^{out} ]=0$, and then one can write a universally valid version of the Arthurs-Kelly relation $\bar{\epsilon}(A)\bar{\epsilon}(B)\geq
|\langle [A, B]\rangle|$ by replacing $B^{out}$ with $N^{out}$ in the identity (2). See also eq.(16).
\end{thebibliography}
\end{document}